\documentclass[twocolumn,showpacs,floatfix,prl]{revtex4}
\usepackage{amssymb}
\usepackage{graphicx}
\usepackage{dcolumn}
\usepackage{bm}
\usepackage[usenames]{color}

\begin{document}

\title{New formulation of the Kubo Optical Conductivity: 
a Shortcut to Transport Properties}
\author{G.~De~Filippis$^1$, V.~Cataudella$^1$, A.~de~Candia $^1$, A.~S.~Mishchenko$^{2}$ and
N.~Nagaosa$^{2,3}$}
\affiliation{$^1$SPIN-CNR and Dip. di Fisica - Universit\`{a} di Napoli
Federico II - I-80126 Napoli, Italy \\
$^2$RIKEN Center for Emergent Matter Science (CEMS), Wako, Saitama 351-0198, Japan\\
$^3$Department of Applied Physics, The University of Tokyo,
7-3-1 Hongo, Bunkyo-ku, Tokyo 113-8656, Japan}


\begin{abstract}

The Kubo formula for  
the electrical conductivity is rewritten in terms of a sum of Drude-like 
contributions associated to the exact eigenstates of the interacting system, 
each characterized by its own frequency-dependent relaxation time. 
The structure of the novel and equivalent formulation, 
weighting the contribution from each eigenstate by its Boltzmann occupation factor,
simplifies considerably the access to the static properties (dc conductivity) 
and resolves the long standing difficulties to recover the Boltzmann result for
dc conductivity from the Kubo formula. 
It is shown that the Boltzmann result, containing the correct transport scattering 
time instead of the electron lifetime determined by the Green function, can be recovered 
in problems with elastic and inelastic scattering at the lowest order of 
interaction.

\end{abstract}

\maketitle

\section{Introduction}

The electrical resistivity of a metal coming from the scattering with phonons or impurities is an 
important topic in the condensed matter physics and it has been addressed by using a 
large number of theoretical methods\cite{bol,bol1,kubo,mori}. 
In particular, one of the most powerfull tool for investigating 
the metal transport properties is represented by the Boltzmann equation\cite{bol,bol1}. 
It is derived on the basis of phenomenological assumptions within a semiclassical approach,
and it is mostly suitable for the calculation of the electrical resistivity in the often 
encountered weak coupling regime. Indeed, by indicating with
$\lambda$ a dimensionless parameter characterizing the strength of the
coupling with phonon or impurities, even in the weak coupling limit, 
contrary to many physical properties, the analysis of 
the dc conductivity, $\sigma_{dc}$, is not a trivial problem since 
$\sigma_{dc}$ displays a singularity at $\lambda=0$, i.e.
$\sigma_{dc} \rightarrow \infty$ when $\lambda \rightarrow 0$\cite{mahan}.
In particular $\sigma_{dc}$ can be expanded
in a Laurent series in $\lambda$, near $\lambda=0$, with the lowest order term of the order of
$\lambda^{-2}$. Although very good for the description of the transport properties at 
small values of $\lambda$, the Boltzmann 
approach can not be systematically extended to any coupling and finite frequencies. 
On the other hand, the dynamic charge response to an electric field can be derived by 
using the quantum linear 
response theory and the Kubo formula\cite{kubo} whose validity 
is not restricted to the weak coupling regime.
However, in the standard Kubo formulation (SKF), it is not 
straightforward to extract the leading
term in the weak coupling limit since low-frequency divergences appear.   
Two remedies to this problem have been proposed in literature.
One is the van Hove's $\lambda^2 t$ limit\cite{lambda2_1,lambda2_2,lambda2_3}, where 
if the limits $\lambda \rightarrow 0$ and $t \rightarrow \infty$ ($t$ is the time)
with $\lambda^2 t=$ const are performed, one gets
an expansion of the dc conductivity where each term is
finite. However, the ad hoc recipe to fix $\lambda^2 t$ 
is not justified. The other proposal proceeds by expressing the response function in terms 
of a self-energy. It is based on the projection technique
introduced by Mori\cite{mori} and Zwanzig\cite{zwan} 
and the memory function formalism\cite{wolfle,kenkre,argyres}. 
In the following we will call it standard formulation of
the optical conductivity (SFOC). 
In this approach, to circumvent the divergence of $\sigma_{dc}$, the idea is 
to expand $1/\sigma_{dc}$ in successive powers of $\lambda$.
Evaluation of the memory function at the lowest order of $\lambda$ gives the classical Drude 
formula $\sigma_{dc}= n e^2 \tau/m$ which, however, contains a relaxation time different from that 
entering into the Boltzmann solution\cite{argyres}.
The last flaw can be fixed, though it requires a trick similar to the joint $\lambda^2 t$ limit: 
within SFOC the correct weak couping limit requires again partial summation of 
infinite series of contributions\cite{argyres}.
 
In this paper we derive a Boltzmann weighted formulation of the optical conductivity (BWFOC), which is
equivalent to the Kubo formula\cite{kubo}, but that has significant advantages over both
Boltzmann solution and SFOC.
BWFOC trivially reproduces the Boltzmann approach results without any artificial conditions of 
joint limits and without the necessity of partial summations of infinite series of contributions. 
On the other hand BWFOC retains all advantages of SFOC, like the 
possibility to consider finite frequencies
and to make a systematic improvement of the result in higher orders of the interaction 
$\lambda$.   

\section{Kubo formula}

The SKF provides the linear response to a small electric field, along $x$ axis, 
of a system in thermodynamic equilibrium (units are such that
$\hbar=1$):
\begin{eqnarray}
\sigma(z)=\frac{i}{z V}\left(\Pi(z)-q_e^2 \Gamma \right),
\label{sigma}
\end{eqnarray}
where $V$ is the system volume, $z$ lies in the complex upper half-plane, $z=\omega+i\epsilon$ with 
$\epsilon>0$, $q_e$ is the electronic charge, the quantity $\Gamma$, in absence 
of superconductivity and in the thermodynamic limit, is given by:
\begin{eqnarray}
q_e^2 \Gamma= -\int_0^{\beta} d\tau  \left\langle J(\tau)J(0) \right\rangle,
\label{gam}
\end{eqnarray}  
and $\Pi(z)$ is the current-current 
correlation function
\begin{eqnarray}
\Pi(z)=-i \int_0^{\infty} dt e^{i z t} \left\langle [J(t),J(0)] \right\rangle.
\label{pi}
\end{eqnarray}
In Eq.~\ref{pi} (Eq.~\ref{gam}) $J(t)$ ($J(\tau)$) is the (imaginary time) 
Heisenberg representation of the 
current operator along the $x$ axis, $[,]$ denotes the commutator, and 
$\left\langle \right\rangle$ indicates the thermodynamical average. 

By choosing the eigenbasis of the interacting system Hamiltonian,
it is straightforward to show\cite{shastry} that the real part of the optical conductvity, 
after performing the limit $\epsilon \rightarrow 0^{+}$, can be written as
\begin{eqnarray}
\Re \sigma(\omega)=D \delta(\omega)+ \sigma_{reg}(\omega),
\label{realsigma} 
\end{eqnarray}
where the regular part $\sigma_{reg}(\omega)$ is defined by:
\begin{eqnarray}
\sigma_{reg}(\omega)= 
\mathop {\sum_{n}\sum_{m}}_{\epsilon_n \neq \epsilon_m}  \frac{\pi}{V} \frac{
| \left\langle \psi_n \right | J
\left | \psi_m \right\rangle |^2} {\omega_{nm}} 
\delta \left( \omega-\omega_{nm} \right ) \left( p_n-p_m\right) \; .\nonumber
\end{eqnarray}
Above $p_n=e^{-\beta \epsilon_n}/Z$ is the Boltzmann weight of the eigenstate 
$  \left | \psi_n \right\rangle$, 
$\epsilon_n$ is the corresponding energy, $Z$ is the partition function, 
$\omega_{nm}=\epsilon_m-\epsilon_n$, $\beta=1/K_{B}T$, $K_{B}$ being the Boltzmann constant, 
and the Drude weight 
$D$, i.e. the coefficient of the zero frequency delta function contribution, is 
given by\cite{notenote} 
\begin{eqnarray}
D= \frac {\pi \beta}{V}  \mathop{\sum_{n}\sum_{m}}_{\epsilon_n=\epsilon_m} p_n
\left | \left\langle \psi_n \right | J
\left | \psi_m \right\rangle \right |^2.
\label{drude}
\end{eqnarray}
$\sigma(\omega)$ satisfies the sum rule\cite{maldague}
\begin{eqnarray}
\int_{-\infty}^{\infty} d\omega \Re \sigma(\omega)=- \frac {\pi q_e^2 \Gamma} {V}.
\end{eqnarray}

The SKF is the most frequently used formulation for the calculation of the quantum 
optical conductivity. However we note that in this formulation 
$\Re \sigma(\omega)$ shows a singularity at 
$\omega=0$ if one proceeds perturbatively. Indeed, at $\lambda=0$, $\sigma_{reg}(\omega)=0$ so that 
only the coefficient $D$ turns to be nonzero. As consequence 
the evaluation of the current-current correlation function by an expansion in a small 
parameter fails due to the singular behavior at small frequencies.

\section{Memory function formulation}

To overcome the difficulties related to 
the diagrammatic techniques that have to deal with summing divergent series, the SFOC was suggested,  
where one represents $\sigma(z)$ in terms of a memory function $M(z)$\cite{wolfle,kenkre,argyres}:
\begin{eqnarray}
\sigma(z)=-\frac {i}{V} \frac{q_e^2 \Gamma}{z+i M(z)},
\label{memory}
\end{eqnarray}
with 
\begin{eqnarray}
M(z)=i \frac{z \Pi(z)}{\Pi(z)-q_e^2 \Gamma}.
\label{memory1}
\end{eqnarray}
This approach, introduced earlier by Kadanoff and Martin\cite{kadanoff}, allows to extract easily
the resonance structures of the optical absorption due to the relaxation processes, 
since the memory function $M(z)$ has a simple expansion in the lowest order
in the impurity concentration and the electron-phonon coupling\cite{wolfle}. Indeed, by taking 
into account that $\Pi(z)$ decreases as $1/z^2$ when $z \rightarrow \infty$, the first step 
is to expand $M(z)$ at high frequencies (short time expansion) so that 
$M(z) \simeq -i z \Pi(z)/q_e^2 \Gamma$. Successively, by using the equations of 
motion of the Green functions, one can express the product $z\Pi(z)$ in terms of 
the force-force correlation function $F(z)$, a Green function, 
involving the commutator between the current operator and the Hamiltonian:
\begin{eqnarray}
z\Pi(z)= \frac {F(z)-F(z=0)}{z},
\label{memory2}
\end{eqnarray}
with
\begin{eqnarray}
F(z)=i \int_0^{\infty} dt e^{i z t} \left\langle [J(t),H],[J(0),H] \right\rangle.
\label{memory3}
\end{eqnarray}

Weak coupling and low frequency limit of SFOC give the classical Drude formula but with a wrong 
relaxation time. 
The relaxation time in the Boltzmann expression is the average of the relaxation times related to the 
eigenstates of the system in absence of the interaction $\left\langle \tau \right\rangle_{av}$. 
On the other hand in SFOC it is $\left ( \left\langle 1/\tau \right\rangle_{av} \right)^{-1}$, i.e. 
since SFOC approach, at the lowest order, averages the inverse relaxation times,  
recovery of the Boltzmann formula requires a procedure 
equivalent to the $\lambda^2 t$ limit.     

\section{New formulation of the optical conductivity}

Here we derive the BWFOC, which overcomes the above described difficulties.   
We note that  $\Pi(z)$ is analytic in the upper half of the complex plane 
and vanishes as $z \rightarrow \infty$. Consequently 
$\Pi(z)$ can be represented as a spectral integral
\begin{eqnarray}
\Pi(z)=  \frac  {1}{\pi} \int_{-\infty}^{\infty} d\omega \frac {\Im{\Pi(\omega)}}{\omega-z}.
\label{spectral}
\end{eqnarray} 

On the other hand $\Im{\Pi(\omega)}$ can be expressed in terms of $\Psi(z)$, 
the Fourier transform of 
symmetrized correlation function $\left\langle (J(t)J(0)+J(0)J(t)) \right\rangle$:
\begin{eqnarray}
\Psi(z)=-i \int_0^{\infty} dt e^{i z t} \left\langle (J(t)J(0)+J(0)J(t)) \right\rangle,
\label{spectral1}
\end{eqnarray}  
i.e. $\Im{\Pi(\omega)}=\tanh(\beta\omega/2)\Im{\Psi(\omega)}$\cite{notenote2}. 
Successively, by introducing the Lehmann representation 
of the correlation function $\Im{\Psi(\omega)}$, using the Eq.~\ref{spectral}, 
and writing the quantity $\Gamma$ in the 
eigenbasis of the interacting system Hamiltonian, one obtains $\Gamma=\sum_n p_n 
\left ( \gamma_n+ \nu_n \right )$ and 
$\Pi(z)=\sum_n p_n \Pi_n(z)$, where:
\begin{eqnarray}
\gamma_n= -\mathop{\sum_{m}}_{\epsilon_n \ne \epsilon_m}
\frac {2 \left | \left\langle \psi_n \right | J
\left | \psi_m \right\rangle \right |^2}{q_e^2 \omega_{nm}}
\tanh(\frac {\beta \omega_{nm}}{2}),
\label{decomposition}
\end{eqnarray}
\begin{eqnarray}
\nu_n=
-\frac {\beta} {q_e^2 } \mathop{\sum_{m}}_{\epsilon_n=\epsilon_m}
\left | \left\langle \psi_n \right | J
\left | \psi_m \right\rangle \right |^2,
\label{decomposition1}
\end{eqnarray}
and   
\begin{eqnarray}
\Pi_n(z)= \mathop{\sum_{m}}_{\epsilon_n \ne \epsilon_m}
\left | \left\langle \psi_n \right | J
\left | \psi_m \right\rangle \right |^2
\tanh(\frac {\beta \omega_{nm}}{2}) f_{nm}^{(a)}(z).
\label{decomposition2}
\end{eqnarray}
Here $f_{nm}^{(a)}(z)= \frac {1} {z-\omega_{nm}}-\frac {1} {z+\omega_{nm}}$.
In terms of the microcanonical quantities $s_n=\gamma_n+ \nu_n$ and $\Pi_n(z)$, the BWFOC reads
\begin{equation}
\sigma(z)= \sum_{n} p_n \sigma_{n}(z) \; ,
\label{main} 
\end{equation}
where
\begin{eqnarray}
\sigma_{n}(z)=  \frac{i}{z V} \left(\Pi_n(z)-q_e^2 s_n \right).
\label{decompositionsigma}
\end{eqnarray}
One can introduce now, for each of the quantum numbers $n$ labelling the eigenstates of the 
Hamiltonian, separate relaxation or memory function $M_n(z)$ (see Appendix for proof) 
\begin{eqnarray}
\sigma_{n}(z)= -\frac{i}{V} \frac{q_e^2 s_n} {z + i M_n(z)}, 
\label{memoryn}
\end{eqnarray}
with 
\begin{eqnarray}
M_n(z)=i \frac{z \Pi_n(z)}{\Pi_n(z)-q_e^2 s_n}.
\label{memoryn1}
\end{eqnarray}
Finally, by taking into account that $z f_{nm}^{(a)}(z)=\omega_{nm} f_{nm}^{(s)}(z)$, where 
$f_{nm}^{(s)}(z)= \frac {1} {z-\omega_{nm}}+\frac {1} {z+\omega_{nm}}$, one can 
express the product $z \Pi_n(z)=f_n(z)$ in terms of the commutator between the current and 
Hamiltonian operators:
\begin{eqnarray}
f_n(z)=\mathop{\sum_{m}}_{\epsilon_n \ne \epsilon_m}
\frac {\left | \left\langle \psi_n \right | [J,H]
\left | \psi_m \right\rangle \right |^2} {\omega_{nm}}
\tanh(\frac {\beta \omega_{nm}}{2}) f_{nm}^{(s)}(z)
\label{force-force}
\end{eqnarray}
that is the analogous of the introduction of the force-force correlation function in the new 
fromulation. The set of equations [15-20] represents the BWFOC. 

The BWFOC restores the semiclassical Boltzmann result at the lowest order in the coupling strength 
but it allows also a non trivial generalization to all frequencies and couplings. 
Namely, in all Boltzmann-like treatments 
a similar formula can be derived  
but with frequency independent memory function  $M_n(z)=1/\tau_n$\cite{devreese,vliet}. 
Furthermore the quantities $s_n$, $\tau_n$, and $p_n$ are exact in BWFOC, 
whereas they are calculated in a perturbative way 
within the Boltzmann approach.

We also point out that SKF and SFOC result in general expressions involving only
the response function which can be represented in any basis. On the other hand, 
the new formulation explicitly relies on the use of eigenstates as basis. 
This more limited choice allows to incorporate explicitly the Boltzmann weight.

In order to recover the Boltzmann result we decompose 
the full Hamiltonian $H$ as $H=H_0+V$, where 
$V$ is the interaction potential which gives rise to dissipation, and suppose that V is 
momentum independent and that the solid is homogeneous. In this case the 
conductivity tensor reduces to just the diagonal terms and they are equal, so that: 
$\sigma_{n}(z)=\sum_{l=1}^d \sigma_{n,l}(z)/d$, where $d$ is the system dimensionality and 
$l$ indicate the lattice axes directions. The 
Eq.~\ref{memoryn} assumes the following form:
\begin{eqnarray}
\sigma_{n}(z)= -\frac{i}{dV} \frac{q_e^2 \bar{s}_n} {z + i \bar{M}_n(z)},
\label{memorynisotropic}
\end{eqnarray}
where $\bar{s}_n=\sum_{l=1}^d s_{n,l}$, $\bar{\Pi}_n(z)=\sum_{l=1}^d \Pi_{n,l}(z)$, and 
$\bar{M}_n(z)=i z \bar{\Pi}_n(z)/(\bar{\Pi}_n(z)-q_e^2 \bar{s}_n)$.  
By approximating 
the exact eigenstates and eigenvalues with that ones of $H_0$, noticing that the matrix elements 
of the current operator between eigenstates of $H_0$ associated to different eigenvalues are zero,  
putting $z=i \epsilon$ and performing the limit $\epsilon \rightarrow 0^{+}$, one obtains:
\begin{eqnarray}
\sigma_{dc}^{(0)}= \frac {\beta} {d V}  \sum_{n} p_n^{(0)} \tau_n^{(0)}
\mathop{\sum}_{\epsilon_n^{(0)}=\epsilon_m^{(0)}}
\sum_{l=1}^d 
\left | \left\langle \psi_n^{(0)} \right | J_l
\left | \psi_m^{(0)} \right\rangle \right |^2 
\label{dc}
\end{eqnarray}
where the relaxation time associated to the eigenstate of $H_0$ with eigenvalue $\epsilon_n^0$ is:
\begin{eqnarray}
\frac {1} {\tau_n^{(0)}}=\pi \frac {\sum_{m,l}  
\left | \left\langle \psi_n^{(0)} \right | [J_l,V]
\left | \psi_m^{(0)} \right\rangle \right |^2 \delta(\epsilon_n^{(0)}-\epsilon_m^{(0)})}
{\mathop{\sum}_{\epsilon_n^{(0)}=\epsilon_m^{(0)}} \sum_{l=1}^d
\left | \left\langle \psi_n^{(0)} \right | J_l
\left | \psi_m^{(0)} \right\rangle \right |^2 },
\label{dc1}
\end{eqnarray}
$J_l$ being the component of the current operator along the l-direction. In the following we 
show that, on the basis of this new formula, some known results can be easily reproduced, but also 
that new results can be deduced in inelastic scattering problems. 

\section{Scattering by impurities}

As first example we consider a noninteracting electron gas scattered 
by spin-independent impurity potentials. In this case 
$H_0=\sum_{\vec{k}} \epsilon_k^{(0)}  c^{\dagger}_{\vec{k}}c_{\vec{k}}$ with 
$\epsilon_k^{(0)}=k^2/2m$ and $J_l=q_e \sum_{\vec{k}} \frac {k_l} {m}  
c^{\dagger}_{\vec{k}}c_{\vec{k}}$. Taking into account that $[J_l,H_0]=0$ and that 
the eigenvectors of the non interacting Hamiltonian 
are labelled by the total wavenumber $\vec{k}$, the matrix element
$\left\langle \vec{k}\right | [J_l,V] \left | \vec{k}^{'} 
\right\rangle$ provides: 
$q_e (k_l-k_{l}^{'}) \left\langle \vec{k} \right | V \left | 
\vec{k}^{'} \right\rangle /m$. 
It is straightforward to show that the dc conductivity becomes:
\begin{eqnarray}
\sigma_{dc}^{(0)}= -\frac {q_e^2} {d V m^2}  \sum_{\vec{k}} f_k^{'} k^2 \tau_k^{(0)},
\label{dcimpurity}
\end{eqnarray}
with 
\begin{eqnarray}
\frac {1} {\tau_k^{(0)}}= 2 \pi \sum_{\vec{k}^{'}}
\left | V_{\vec{k},\vec{k^{'}}}  \right|^2
\delta(\epsilon_k^{(0)}-\epsilon_{k^{'}}^{(0)}) 
(1-\cos(\theta_{\vec{k},\vec{k}^{'}})). 
\label{tauk}
\end{eqnarray}
Here $\theta_{\vec{k},\vec{k}^{'}}$ denotes the angle between $\vec{k}$ and $\vec{k}^{'}$, and 
$f_k^{'}$ represents the derivative of the Fermi distribution 
with respect to the energy $\epsilon_k^{(0)}$. The set of 
Eq.~\ref{dcimpurity} and Eq.~\ref{tauk} 
coincides with the semiclassical result provided by the Boltzmann equation\cite{mahan}. 
In particular the factor 
$1-\cos(\theta_{\vec{k},\vec{k}^{'}})$ shows that Eq.~\ref{tauk} represents the correct
transport scattering time. 

\section{Inelastic scattering: the Fr\"{o}hlich polaron}

As second example we consider 
the Fr\"{o}hlich polaron model\cite{Frohlich, defilippis} where electron ($\vec{r}$ and $\vec{p}$
are the position and momentum operators) is scattered by phonons
($a^{\dagger}_{\vec{q}}$ the creation operator with wave number $\vec{q}$)
with interaction vertex $M_q=i \omega_0 \left( R_p 4 \pi
\alpha/q^2 V \right) ^{1/2}$:
\begin{equation}
H=p^2/2m + \omega_0 \sum_{\vec{q}}a^{\dagger}_
{\vec{q}}a_{\vec{q}} + \sum_{\vec{q}} [M_q e^{i \vec{q} \cdot \vec{r}  }
a_{\vec{q}}+h.c.] .
\end{equation}
Here $\alpha$ is the dimensionless coupling constant,
$R_p= \left( 1/2m\omega_0 \right) ^{1/2}$,
and $V$ is the volume of the system. 

Due to the inelastic nature of the scattering processes, the theoretical treatment is 
complicated\cite{mobf,devreese} and different approaches give different expressions even in
the limit of very low temperature. These various methods usually agree in the weak 
coupling limit ($\alpha \ll  1$) providing for the mobility ($\mu=\sigma_{dc}/n q_e$, where $n$ is 
the particle density)\cite{mahan}: 
\begin{equation}
\mu=\frac {q_e}{2 \alpha m \omega_0} N_0.
\label{mobilityf}
\end{equation}
Here $N_0=1/(e^{\beta \omega_0}-1)$ is the phonon number density.

This result can be derived from the Kubo formula\cite{mahan}. 
The first term of the expansion 
of the S matrix leads to the bubble diagram including two electronic Green functions 
$G(k,\omega)$, which, in turn, are obtained by Dyson's equation at the lowest order in 
the electron-phonon coupling $\alpha$. This procedure leads to $\mu=q_e \tau /m$, where 
$\tau=1/2\alpha N_0 \omega_0$ and then Eq.~\ref{mobilityf} is recovered. 
However, in this approach, 
$\tau$ coincides with the electron lifetime derived from the Green function
$G(k=0)$ and does not include 
the equivalent of the $1-cos(\theta_{\vec{k},\vec{k}^{'}})$ factor in the elastic scattering. 
On the other hand the Drude formula involves the transport scattering time, related to
the real part of the memory function, which, in general, is not identical with the single-particle
scattering time, that is related to the imaginary part of the self-energy of the electron propagator. 

Another approach to derive the polaron mobility in the weak coupling limit is 
based on the Boltzmann equation. By neglecting the {\it in}-scattering terms contribution in the 
collision term\cite{kadanoff1}, one obtains again Eq.~\ref{mobilityf}. 
It turns out that Eq.~\ref{mobilityf} does not agree with correct solution of the Boltzmann 
equation in the relaxation time approximation (see discussion by Sels and Brosens\cite{brosens}). 

The path integrals method adds a result in disagreement with the other approaches.
In the low temperature and weak coupling limits, the polaron mobility in 
Feynman-Hellwarth-Iddings-Platzman (FHIP)\cite{feynman} approach
differs from Eq.~(\ref{mobilityf}) by a factor of $3 K_B T/2 \omega_0$.
It has been shown that the result obtained in Ref.~\cite{feynman} can be obtained by using the 
memory function formalism and the Feynman polaron model\cite{devreese1}, so that 
the mobility, in this approach, suffers the problem related 
to the average value of $1/\tau$ rather than $\tau$. 

BWFOC allows trivial derivation of the correct perturbative solution of the polaron mobility. 
By taking 
into account that $J_l=q_e p_l/m$ and $[p_l,V]= 
\sum_{\vec{q}} q_l [M_q e^{i \vec{q} \cdot \vec{r}} a_{\vec{q}}-h.c.]$, from Eq.~\ref{dc1} 
one obtains the relaxation time $1/\tau_k^{(0)}=1/\tau_{a,k}^{(0)}+1/\tau_{e,k}^{(0)}$, where 
$1/\tau_{a,k}^{(0)}$ and $1/\tau_{e,k}^{(0)}$ denote 
the contributions coming from absorption and emission 
of longitudinal otptical phonons respectively:
\begin{equation}
1/\tau_{a,k}^{(0)}=\pi \sum_{\vec{q}} \frac {q^2}{k^2} \left | M_q \right | ^2 N_0 
\delta (\epsilon_{\vec{k}}^{(0)}-\epsilon_{\vec{k}+\vec{q}}^{(0)}+\omega_0)
\label{mobilityfa}
\end{equation}
and 
\begin{equation}
1/\tau_{e,k}^{(0)}=\pi \sum_{\vec{q}} \frac {q^2}{k^2} \left | M_q \right | ^2 \left ( 
1+N_0 \right)
\delta (\epsilon_{\vec{k}}^{(0)}-\epsilon_{\vec{k}-\vec{q}}^{(0)}-\omega_0).
\label{mobilityfe}
\end{equation}
We emphasize that the factor $q^2/k^2$, where $\vec{q}$ is the transferred momentum by phonons
in the scattering, is a substitute of the factore $2(1-cos(\theta_{\vec{k},\vec{k}^{'}}))$. 
Hence, BWFOC automatically introduces transport scattering time into perturbative expressions. 
It is remarkable that this factor, introduced phenomenologically
by Fr\"{o}hlich in 1937\cite{frohlich1,frohlich2}, had been discarded in all successive treatments
but has been put back by perturbative expansion of BWFOC.   
Furthermore, 
at low temperatures, where only momenta around $k=0$ contribute to the mobility, one obtains
$\tau_{k}^{(0)} \simeq \tau k^2/ m\omega_0 $, i.e., as it is expected, 
the transport relaxation time $\tau_{k}^{(0)}$ differs by a factor
$ k^2/ m\omega_0$ from the single particle scattering time $\tau$. Finally,
we note that in BWFOC expansion at low temperatures
only the phonon absorption processes contribute to the mobility, that reflects the impossibility
of the events in which a low energy polaron emits a phonon\cite{notenote3}. 

By inserting the time relaxation expression in Eq.~\ref{dc} we obtain the mobility 
in the weak coupling regime at low temperatures as $\mu=\mu_{FHIP} 10/3$, i.e. the mobility 
differs by a numerical factor $10/3$ from the result of FHIP\cite{feynman}
and by $5 k_B T/\omega_0$ from the value obtained through 
the diagrammatic technique\cite{finalnote}, i.e. Eq.~\ref{mobilityf}. 

In this paper we derived a new formulation of the optical conductivity which allows 
a trivial derivation of the Boltzmann result. The structure of BWFOC (\ref{main}), weighting 
the contribution from exact eigenstates by Boltzmann occupation numbers, 
allows to treat weak coupling and low temperature limits trivially, which
is in complete contrast with all previous formulations of the optical conductivity. 
Beyond recovery of the correct Boltzmann limit, BWFOC retains 
possibility to consider finite frequency features and perform calculations 
in the intermediate and strong coupling regimes. We demonstrated the 
power of BWFOC formulation for elastic and inelastic scattering problems. 

\appendix*

\section {Appendix}

The new formulation of the linear response theory is based on the idea to
introduce, for each of the quantum numbers $n$ labelling the eigenstates of the
Hamiltonian, one relaxation or memory function $M_n(z)$:
\setcounter{equation}{0}
\renewcommand{\theequation}{A.\arabic{equation}}
\begin{eqnarray}
\sigma(z)= -\frac{i}{V} \sum_{n} p_n  \frac{q_e^2 s_n} {z + i M_n(z)},
\label{memoryna}
\end{eqnarray}
with
\begin{eqnarray}
M_n(z)=i \frac{z \Pi_n(z)}{\Pi_n(z)-q_e^2 s_n}.
\label{memoryn1a}
\end{eqnarray}

Here we want to prove that the quantity $\Pi_n(z)-q_e^2 s_n$ is different from zero for $\Im {z}
\ne 0$. We observe that by using the spectral representation
\begin{eqnarray}
\Pi_n(z)=  \frac  {1}{\pi} \int_{-\infty}^{\infty} d\omega \frac {\Im{\Pi_n(\omega)}}{\omega-z}.
\label{spectrala}
\end{eqnarray}
we have, for $z=x+i \epsilon$:
\begin{eqnarray}
\Pi_n(x+i \epsilon)=
\frac  {1}{\pi} \int_{-\infty}^{\infty} d\omega \frac {\Im{\Pi_n(\omega)}
(\omega-x+i\epsilon)}
{(\omega-x)^2+\epsilon^2}.
\label{spectral1a}
\end{eqnarray}
Since $s_n$ is real, first of all
we find the values $x+i \epsilon$, with $\epsilon \ne 0$, for which $\Pi_n(z)$ is real.
For these values we have:
\begin{eqnarray}
\int_{-\infty}^{\infty} d\omega \frac {\Im{\Pi_n(\omega)}}
{(\omega-x)^2+\epsilon^2}=0.
\label{spectral2a}
\end{eqnarray}
Next step is to write the denominator $D_n(z)$ of $M_n(z)$
in the complex upper half-plane for $z$ values
where Eq.~\ref{spectral2a} is satisfied:
\begin{eqnarray}
D_n(x(\epsilon)+i\epsilon)=
- \int_{-\infty}^{\infty} d\omega \frac {\Im{\Pi_n(\omega)}} {\pi \omega}
\frac {x^2+\epsilon^2} {(\omega-x)^2+\epsilon^2} -q_e^2 \nu_n > 0,
\nonumber
\end{eqnarray}
having taken into account that $s_n=\gamma_n+ \nu_n$, $\gamma_n=\Pi_n(z=0)/q_e^2$,
$-\Im{\Pi_n(\omega)}/\omega \ge 0$ and $\nu_n \le 0$.
This proves that $M_n(z)$ is analytic in the complex upper half-plane. A similar proof
has been given\cite{gotze} to justify the introduction of the memory function in the Eq.~8
of the main text.


\begin{thebibliography}{99}
\bibitem{notenote} In particular it has been conjectured\cite{zotos} that: i) integrable systems
exhibit ballistic behaviour, i.e. dissipationless finite temperature conductivity,
if the Drude coefficient is different from zero at $T=0$, and remain ideal insulators, 
i.e. $D(T>0)=0$, at all temperatures if $D(T=0)=0$; ii) a generic nonintegrable system
displays finite conductivity at finite temperature, i.e it is characterized by diffusive motion,
being $D(T>0)=0$. We will restrict our attention to normal conductors, i.e. 
$\lim_{\epsilon \to 0^{+}} (\Pi(i \epsilon)-q_e^2 \Gamma)=D(T>0)=0$. 
On the other hand we emphasize that in these systems $D(T>0)=0$ only if the exact eigenstates 
are used in the Eq.~\ref{drude}, so that we will retain this contribution up to the end of the 
calculation.
\bibitem{frohlich2} H. Fr\"{o}hlich and N. F. Mott, Proc. R. Soc. Lond. A {\bf 171}, 496 (1939).
We note that, within this reference, the equations determining the transport scattering time
coincide with Eq.~\ref{mobilityfa} and Eq.~\ref{mobilityfe} except for the factor
$q^2/k^2$ that is replaced with $-2q_x/k_x$. Fr\"{o}hlich and Mott themselves demonstrated, 
by using the energy conservation, that the factor 
$-2q_x/k_x$ is equivalent to the quantity $(q^2 \pm q_p^2)/k^2$ ($+$ ($-$) 
for one phonon emission (absorption)), 
where $q_p^2/2m=\omega_0$. 
On the other hand, the inspection of the Boltzmann equation reveals that the contribution 
$-2q_x/k_x$ is correct 
only in the elastic scattering problems (where $q_p=0$ and, in this case, 
all the approaches provide the same result). 
We also note that the term $q_p^2/k^2$ has been neglected by Frohlich himself 
in ref.~\cite{frohlich1}.
Actually it is not clear which is the 
appropriate factor in the inelastic scattering problems, and, in general, 
it is discarded (in this way 
transport scattering time and electron lifetime coincide). 
Then the problem is open, and, so far, not solved yet. 
This does not allow to establish which is 
the correct perturbative limit for the mobility. 
Within BWFOC the appropriate factor 
both in elastic and inelastic scattering problems turns out to be $q^2/k^2$. 
We note, in particular, that the validity of our approach is not related to 
the nature of the scattering events.
\bibitem{notenote3} Note the strange fact that in the FHIP\cite{feynman} 
treatment mobility is dominated by phonon emission processes\cite{devreese}.
\bibitem{finalnote}  We emphasize that, within the SKR,
the second vertex correction provides a further factor in Eq.~\ref{mobilityf}
of the order of $k_B T/\omega_0$\cite{mahan}, in agreement, apart from a numerical factor,
with the prediction of WBFOC (we note that WBFOC obtains this result at the lowest order).
\end{thebibliography}
\end{document}